\documentclass[11pt,aps,groupedaddress,preprintnumbers,
               nofootinbib]{revtex4}

\usepackage{graphicx}  
\usepackage{epsfig}    

\def\rmp#1#2#3{{\it Rev.\ Mod.\ Phys.} {\bf #1}, #2 (#3)}
\def\plb#1#2#3{{\it Phys.\ Lett.} {\bf B#1}, #2 (#3)}
\def\prd#1#2#3{{\it Phys.\ Rev.} {\bf D#1}, #2 (#3)}

\def\prl#1#2#3{{\it Phys.\ Rev.\ Lett.} {\bf #1}, #2 (#3)}

\def\sss{\scriptscriptstyle}	

\def\thingie{\hbox{\kern-9pt\raise1pt%
         \hbox{{\fiverm(}{\lower1.5pt\hbox{\twelvebf--}}{\fiverm)}}}}
\def\pmdiff#1#2{\raise.5ex\hbox{$\sss +#1}$%
    \kern-2.8em\lower1ex\hbox{${\sss-#2}$}} 

\def\barp{{\raise.35ex\hbox{${\sss (}$}}---{\raise.35ex\hbox{${\sss )}$}}}				
\def\bdbarp{\hbox{$B_d$\kern-1.4em\raise1.4ex\hbox{\barp}}}
\def\nlpbarp{\hbox{$\nu_{\ell^{\prime}}$\kern-1.4em \raise1.4ex\hbox{\barp}}}

\def\dm2{\delta M^2_{ m \, m^{\sss\prime}}}

\def\decayarrow{\kern0.2em\hbox{$\raise1.08ex\hbox{\big|}\kern-0.5em
                \longrightarrow$}}

\def\ra{\rightarrow}

\newcommand{\beq}{\begin{equation}}
\newcommand{\eeq}{\end{equation}}

\begin{document}

\title{A Scalar Doublet at the Tevatron?}
\author{Gino Segr\`{e}}
\email{segre@dept.physics.upenn.edu}
\affiliation{\it Department of Physics \& Astronomy, University of Pennsylvania, Philadelphia, PA 19104 USA}
\author{Boris Kayser}
\email{boris@fnal.gov}
\affiliation{\it Theoretical Physics Department, Fermilab, P.O. Box 500, Batavia, IL 60510  USA}
\date{\today}

\preprint{FERMILAB-PUB-11-211-T}

\begin{abstract}
We propose a particularly simple explanation of the W+dijet excess reported by the CDF collaboration. No symmetries beyond those of the standard model are necessary, and the only new particles involved are a spin-zero weak SU(2) doublet with no vacuum expectation value. Possible tests of the model and comparisons to other proposals are discussed.
\end{abstract}

\maketitle


A great deal of excitement has followed the recent announcement from the CDF collaboration \cite{ref1}, \cite{ref2} of an unexpected peak in $W$+ dijet production. The manifestation is an excess in the distribution of events when their number is plotted versus the invariant mass of the dijets. This anomaly appears to indicate the presence of at least one new particle in the 150 GeV range; moreover, such a particle needs to have novel features in both its production and its decay. In this note we propose a particularly simple explanation for this data set. It requires nothing more than the existence of a single SU(2) doublet spin-zero field, similar  to a conventional Higgs field, but differing from it by having no vacuum expectation value (vev). 

The other explanations that have been proposed in recent weeks to explain the $W$+ dijet events have fallen largely, as emphasized by Buckley{\it  et al.} \cite{ref3}, into two classes, both with long histories of consideration. In the first class of models \cite{ref3}-\cite{ref10}, a new gauge boson, a $Z^\prime$, is produced in association with the $W$. In order to explain the data, this new gauge boson's leptonic decay modes need to be suppressed, or, in other words, the $Z^\prime$ has to be leptophobic. This might seem strange, were it not that additional $Z$ bosons of this type are well motivated by extensions of the standard model in which baryon number is promoted from a global to a gauged symmetry \cite{ref10a}; these models also have additional features that make them attractive \cite{ref10b}-\cite{ref14}, e.g., by providing dark matter candidates.

The second class of models seeking to explain the recent data makes use of multiple new particles. These models, as is the case of leptophobic $Z^ \prime$s, are motivated by the possible existence of new symmetries in nature. One interesting proposal involves supersymmetry with or without broken R-parity \cite{ref15}-\cite{ref16}. Another has been provided by Eichten {\it et al.} \cite{ref17}. It makes use of technicolor, an alternative to employing the Higgs mechanism in order to generate masses. Eichten {\it et al.}, long-time students of technicolor (\cite{ref17}, \cite{ref18} and references therein), suggest that the production of a technirho followed by its decay to a $W$ plus a technipion in the 150 GeV mass range is a possible explanation of the new data. Another proposal \cite{ref19} seeks to introduce new particles that may be involved in $B-\overline{B}$ mixing.

Our attention to the model we suggest as an explanation for the $W$ + dijet data was stimulated by the necessity of such a scalar field in a model we recently proposed for electroweak scale leptogenesis \cite{ref20}. We hasten, however, to add that there is not necessarily a correlation between our proposal for leptogenesis and the present one for $W$ + jets events; the resemblance may simply be accidental. We also note that additional spin-zero particles are present in a number of other phenomenological extensions of the standard model. \cite{ref21}

Although our proposal does not depend directly on the additional presence of conventional Higgs fields, some of its features are most easily described if the latter do exist. Let us assume this to be true and denote by $\phi_a$ the conventional Higgs doublets, where $a = 1,2\ldots$, not specifying for the moment the number of such doublets. Calling $\phi$ the additional scalar doublet we are introducing, the terms in the scalar potential that involve this field can be written as
\begin{eqnarray}
V(\phi) & = & \mu^2\phi^\dagger\cdot\phi + \lambda_\phi(\phi^\dagger\cdot\phi )^2 + 
\sum_{a,b} [\lambda^{(1)}_{ab}(\phi^\dagger\cdot\phi )(\phi^\dagger_a\cdot\phi_b)  \nonumber  \\
	& & +  \; \lambda^{(2)}_{ab}(\phi^\dagger\cdot\phi_a )(\phi^\dagger\cdot\phi_b)
	+ \lambda^{(3)}_{ab}(\phi^\dagger_a\cdot\phi )(\phi^\dagger\cdot\phi_b)] + \mathrm{h.c.}
\label{eq1}
\end{eqnarray}
Note that the $\phi$ field has no vev because the mass term parameter $\mu^2$ is taken to be positive and because we have chosen the potential such that no terms linear in the $\phi$ field are present. This restriction can be relaxed somewhat, but for the moment it is simpler to say such terms are simply absent. The mass term parameter in the scalar potential will be taken, as in our leptogenesis model, to be $\mu \sim$ 300 GeV.

If we select the quartic couplings to be ${\cal O}$(1), we observe that, due to the vevs of the conventional Higgs fields, there is a mass splitting between the charged and neutral components of the $\phi$ field that is $\sim \lambda v$, where $v$ is the electroweak symmetry breaking scale and $\lambda$ is a characteristic value for the quartic couplings. That is,
\beq
M_{\phi^+} - M_{\phi^0} \sim \lambda v \sim 100\,\mathrm{GeV} ~~.
\label{eq2}
\eeq
We have taken $\phi^+$ as the more massive component of the doublet, but depending on the signs of the quartic coupling constants, it could equally well be $\phi^0$.

The $\phi$ fields form an SU(2) doublet with the conventional coupling to the electroweak gauge bosons, but,  since $\phi^0$ has no vev, there is no direct correlation between the strength of the doublet's coupling to quarks and leptons and the masses of the latter. We shall denote such couplings by $y$, realizing of course that a different value of $y$ can be chosen for each fermion coupling, as is true for Higgs fields, but now without the restrictions caused by the generation of fermionic masses. Some bounds on the couplings do need to be present, however, in order to satisfy limits imposed by the apparent absence of flavor changing neutral currents.

Denoting the left-handed quark doublets by $\left( \begin{array}{c} u_i \\ d_i  \end{array} \right) _L, i = 1,2,3 $, we can write a Yukawa coupling of quark fields as
\beq
{\cal L}_{\mathrm{Yuk}} = y_{ij} (\overline{u}_{iL} \,\phi^+ + \overline{d}_{iL}\, \phi^0) d_{jR} + 
	y_{ij}^\prime (\overline{u}_{iL}\, \phi^0 - \overline{d}_{iL}\, \phi^-)u_{jR},\;   i,j = 1,2,3
\label{eq3}
\eeq
with of course a similar coupling of $\phi$ to lepton fields.

If we choose as a generic value $y \sim 10^{-2}$, and the mass difference between $\phi^+$ and $\phi^0$ is large enough for the decay
\beq
\phi^+ \ra \phi^0 + W^+
\label{eq4}
\eeq
to proceed, then this process will dominate the $\phi^+$ decays, since it is governed by the relatively large semiweak coupling $g \simeq 0.65$ of the $W$ boson to scalar doublets. We assume that this process does dominate, in which case the width of $\phi^+$ will be of the order of (1 - 10) GeV. We envision the chain that leads to the W + dijet as
\beq 
\overline{q} q \ra \phi^+ \ra W^+ + \phi^0 \ra W^+ + \mathrm{dijet} ~~.
\label{eq5}
\eeq

Our scheme cleanly predicts several angular distributions. Owing to the spinless nature of $\phi^+$ and $\phi^0$, the angular distribution of the $\phi^0$ and $W^+$ from the decay $\phi^+ \ra \phi^0 + W^+$ must be isotropic in the $\phi^+$ rest frame, and the angular distribution of the two jets from the decay $\phi^0 \ra$ jet + jet must be isotropic in the $\phi^0$ rest frame. By angular momentum conservation, the $W^+$ from the decay $\phi^+ \ra \phi^0 + W^+$ must be longitudinally polarized. Hence, its weak decay will produce quark jets or leptons with a $\sin^2 \theta$ angular distribution in the $W^+$ rest frame. Here, $\theta$ is the outgoing jet or lepton direction relative to the $W$ momentum in the $\phi^+$ rest frame.

Given the number of proposals for explaining the new CDF data, it will be interesting to look for ways to distinguish one from another. We illustrate the distinctions by comparing our model to e.g. the one suggested by Eichten {\it et al.}\cite{ref17}, which we will call ELM.

One striking difference between our model and ELM follows from the technirho and technipion both being isospin triplets. This implies that in ELM, the decay channel  $Z$ + dijets should be comparable in magnitude to $W$ + dijets, whereas in our model it is absent. A second difference follows from the presence in the ELM model of the  technivector mesons $\omega^T(0^-1^{- -})$ and $a^T(1^- 1^{++})$, expected to be close in mass to the technirho. Another clear distinction  between our model and the ELM technicolor proposal lies in the couplings to fermions of the new particles. Whereas technipions are presumed to have larger couplings to more massive fermions, those of the $\phi^0$ particle are arbitrary, as we have already mentioned. There might even be significant couplings to leptons. One might  expect, therefore, that the $W$ + dijet distribution is accompanied by a significant peak in, e.g., $W + \mu^+ \mu^-$, or $W + e^+ e^-$. It would be worthwhile to look for such a correlation.

We need to address the question of detecting the lower mass $\phi^0$ particle in $\overline{q} q$ collisions (this question is of course parallel to that of detecting technipions). Since we have taken the fermion coupling constants to be of order $10^{-2}$, the production and subsequent decay of a $\phi^0$ may be hard to detect against the background.
If the $\phi^0$ is somewhat more massive than 160 GeV, we would expect its leading decay mode to be into a pair of gauge bosons, $W^+ W^-$, but this mode may be  below threshold and hence absent.
If the $\phi^0$ coupling to electrons is appreciable, the possibility of detecting $\phi^0$ as a resonance in an electron-positron collision would of course be particularly interesting.

In conclusion, we note that our model presents a simple explanation of the recent CDF $W$  + dijet data, while suggesting a number of other interesting and testable features. It will be interesting to see what further experimental data will reveal.

\section*{Acknowledgments}

We are indebted to Susan Kayser for technical assistance. We thank Estia Eichten for useful conversations.

\end{document}